\documentclass[
    aps, 
    prb, 
    preprint, 
    superscriptaddress, 
    floatfix, 
    amsmath,
    amssymb
]{revtex4-2}

\usepackage{graphicx}
\usepackage{bm}
\usepackage{orcidlink}
\usepackage{booktabs}
\usepackage{tabularx}
\usepackage{amsmath}
\usepackage{hyperref}

\usepackage[mathlines]{lineno}

\begin{document}


\title{Self-trapped holes and acceptor impurities in orthorhombic $\mathrm{\kappa-Ga_2O_3}$}


\author{Eric Welch\orcidlink{0000-0002-8066-8005}}
\affiliation{Department of Physics and Chemistry, Prairie View A\&M University, Prairie View, TX 77446}
\affiliation{Department of Physics, Wright State University, Dayton, OH 45435}

\author{Nathan Rabelo Martins\orcidlink{0000-0002-2229-5044}}
\affiliation{Department of Physics, Texas State University, San Marcos, TX 77666}
\affiliation{Instituto Federal de Educação, Ciência e Tecnologia de Minas Gerais (IFMG), Arcos, MG 35600-306, Brazil}

\author{Luisa Scolfaro\orcidlink{0000-0002-5529-3213}}
\affiliation{Department of Physics, Texas State University, San Marcos, TX 77666}

\author{Luiz A. F. C. Viana\orcidlink{0000-0001-5475-8077}}
\affiliation{Instituto Federal de Educação, Ciência e Tecnologia de Minas Gerais (IFMG), Arcos, MG 35600-306, Brazil}
\affiliation{Instituto de Ciências Exatas e Tecnologia, Universidade Federal de Viçosa, Rio Paranaíba, MG 38810-100, Brazil}

\author{Pablo D. Borges\orcidlink{0000-0003-3829-5134}}
\affiliation{Department of Materials Science, Military Institute of Engineering - IME. Praça General Tibúrcio 80, Urca, Rio de Janeiro 22290-270, RJ, Brazil}

\begin{abstract}

$\mathrm{\kappa-Ga_2O_3}$ is a metastable polymorph of $\mathrm{Ga_2O_3}$ that has attracted increasing interest due to its unique structural and electronic properties with respect to the well-studied $\beta$ phase. Although hole self-trapping has been established in several $\mathrm{Ga_2O_3}$ polymorphs to include the $\kappa$ phase, the behavior of holes in $\mathrm{\kappa-Ga_2O_3}$ and their interaction with substitutional impurities remain less well understood.  In this work, the interplay between hole localization and cation impurities in $\mathrm{\kappa-Ga_2O_3}$ is investigated using hybrid density functional theory.  The stability of the hole is studied in the pure phase along with Al, In, Mg, and Zn as substitutions at the Ga site.  The results show that $\mathrm{\kappa-Ga_2O_3}$ supports a stable hole polaron localized at an O site, while the presence of cation impurities modifies the stability and character of the localized hole state.  Isoelectronic substitution by Al and In suppresses hole localization, resulting in a larger spread of spin density across the cell, while Mg and Zn preserve oxygen-centered hole states with some density shared with the impurity in the Zn system.  Defect formation energies are also evaluated as a function of the oxygen chemical potential and Fermi energy to assess the thermodynamic stability of these impurities and the role of compensating intrinsic vacancies.  Formation energy calculations reveal that O vacancies are the most favorable defect species and most likely compensate for the effects of extrinsic defects.
\end{abstract}

\maketitle

\section{\label{sec:i}Introduction}
Gallium oxide ($\mathrm{Ga_2O_3}$) is a wide bandgap oxide (WBO) semiconductor with numerous polymorphs possessing distinct structural and electronic properties.  The monoclinic $\beta$ phase is the most thermodynamically stable and has been studied extensively including the formation of polarons and defects.  More recently, metastable polymorphs like the orthorhombic $\kappa$ phase have garnered attention due to a polar crystal structure and anisotropic optoelectronic properties \cite{lyons2019electronic,kim2018first,kim2025implementing}.  These differences can influence charge localization and defect formation relative to the stable $\beta$ phase. 

In WBOs holes tend to localize as small polarons confined within a single unit cell.  In these materials where the valence band is dominated by oxygen 2p states, holes prefer to trap and localize on oxygen atoms rather than remaining delocalized due to the flat, low dispersion valence band.  This self-trapping process is accompanied by local lattice distortions that stabilize the hole.  Such hole polarons have been observed or predicted in a wide range of oxide materials and play a key role in determining charge transport and defect energetics \cite{lyons2019electronic,lyons2022self,varley2012role}.  Because hole polarons modify the local electronic structure and lattice configuration, their formation can modify the energetics of intrinsic and extrinsic defects.

First-principles calculations have shown that self-trapped hole (STH) formation is favored in $\mathrm{Ga_2O_3}$ and that the stability of the localized state depends on the local lattice distortions accompanying the trapping \cite{lyons2019electronic,lyons2022self,varley2012role,gake2019first}.  The formation of a hole polaron can, therefore, be used as a tool to study defect physics in a material, since localized holes will interact with intrinsic and extrinsic defects in the host material.  As a result, the presence of hole polarons can modify defect formation energies and influence charge compensation mechanisms.

Substitutional impurities provide a means of modifying the local chemical environment in which hole polarons form.  Isoelectronic substitutions alter the bonding environment without directly introducing free carriers, whereas acceptor impurities intrinsically introduce defect states that can interact with localized holes.  Alloying and doping with these elements have been investigated in $\mathrm{Ga_2O_3}$ polymorphs to understand their influence on the electronic structure and defect chemistry of the material \cite{seacat2021properties,li2012electronic,zhang2012effects, liao2025exploration, welch2024indium}.  However, the interaction between substitutional impurities and hole polarons in the $\kappa$ phase of $\mathrm{Ga_2O_3}$ has not been systematically explored.

In this work, we investigate hole self-trapping in $\mathrm{\kappa-Ga_2O_3}$ and examine how cation substitutional impurities influence the stability of the localized hole state.  Using hybrid density functional theory, we consider both isoelectronic substitutions ($\mathrm{Al_{Ga}}$ and $\mathrm{In_{Ga}}$) and acceptor impurities ($\mathrm{Mg_{Ga}}$ and $\mathrm{Zn_{Ga}}$) on the Ga site.  The energetics of hole trapping and defect formation are analyzed through calculations of self-trapping energies, lattice distortions associated with polaron formation, and defect formation energies under different chemical potential conditions.  The remainder of this paper is organized as follows: Section II describes the computational methods, Section III presents the calculated results and discussion, and Section IV summarizes the conclusions.

\section{\label{sec:methods}Methodology}
The properties of $\mathrm{\kappa-Ga_2O_3}$, hole localization, and the effects of isoelectronic (Al and In) and acceptor (Mg and Zn) impurities at the cation site on hole trapping are investigated here using hybrid density functional theory.  Hybrid functionals have previously been shown to accurately reproduce experimental lattice parameters \cite{krukau2006influence} and localized defect states \cite{lyons2019electronic,lyons2022self,varley2012role,lyons2018survey} in $\mathrm{Ga_2O_3}$ and other similar WBOs. 

Calculations were performed using the projector augmented wave (PAW) \cite{kresse1999ultrasoft} method as implemented in VASP \cite{kresse1996efficiency,kresse1996efficient,kresse1993ab} with hybrid screened exchange-correlation treated at the HSE06 level and a planewave cut off energy of 500 eV.  To reproduce the experimental band gap of 4.9 eV \cite{banda2024structural}, the fraction of Hartree–Fock exact exchange was set to 0.32 for the $\kappa$ phase and all defect-containing systems. Earlier studies of $\mathrm{\kappa-Ga_2O_3}$ reported smaller band gaps; however, more recent experimental measurements indicate a larger value consistent with the present calculations \cite{banda2024structural}.

Brillouin zone sampling was performed using a 2×2×2 $\Gamma$-centered k-point mesh for all supercell calculations; $\Gamma$-point only calculations were tested and some systems were unable to find electronic/ionic convergence using a single k-point. 80 atom supercells were used for all systems including spin polarization explicitly with symmetry switched off.  The initial magnetic configuration was chosen to favor localization of the hole on a single oxygen atom.  Initial geometries were obtained from Kim et al \cite{kim2018first} and expanded to form supercells in which each lattice vector exceeded 10 Å to minimize interactions between periodic images.  Defect supercells were created using the doped \cite{kavanagh2024doped} and shakenbreak \cite{mosquera2022shakenbreak} python libraries to adequately sample the potential energy surface and find the global ground state, using the standard inputs for bond distortion and rattling ranges.  The lowest energy configuration of each impurity structure was chosen after initial PBEsol ionic relaxation and were further relaxed using HSE06 exchange-correlation and standard down sampling methods, as outlined in the doped documentation.  To assess supercell size effects, 160-atom supercells of $\mathrm{\kappa-Ga_2O_3}$ in both the neutral and self-trapped hole configurations were also tested.   The difference in self-trapped hole energies between the 80- and 160-atom supercells was found to be only 0.04 eV, and the difference in structural distortions surrounding the hole was less than 0.02 Å. These results indicate that the 80-atom supercell provides an adequate description of hole localization in $\mathrm{\kappa-Ga_2O_3}$, as seen in other works \cite{lyons2019electronic}.  Ga 3d electrons were included with the valence electrons to maintain consistency with the pseudopotentials used for impurity atoms with occupied d-orbitals (Zn and In); the valence configurations used were Ga $\mathrm{3d^{10} 4s^2 4p^1}$, O $\mathrm{2s^2 2p^4}$, Al $\mathrm{3s^2 3p^1}$, In $\mathrm{4d^{10} 5s^2 5p^1}$, Mg $\mathrm{3s^2}$, and Zn $\mathrm{3d^{10} 4s^2}$.

Hole localization in $\mathrm{\kappa-Ga_2O_3}$ was first examined in the absence of impurities to determine the intrinsic stability of the STH.  The stability of the localized hole state can be quantified through the self-trapping energy $\mathrm{E_{ST}}$, defined in a similar way to Varley et al \cite{varley2012role},

\begin{equation}
\label{eq1}
\mathrm{E_{ST}={E_{deloc}-E_{STH}-\Delta \bar{V}}}
\end{equation}

where $\mathrm{E_deloc}$ is the energy of the configuration in which the hole exists in the undistorted lattice prior to structural relaxation, $\mathrm{E_{STH}}$ is the energy of the relaxed configuration in which the hole is localized, and $\mathrm{\Delta \bar{V}}$ is the alignment of the average electrostatic potential between the two supercells far away from the polaron site; a positive value indicates favorable hole trapping.  The delocalized reference state was obtained from a single-point calculation in which an electron was removed from the neutral supercell without allowing ionic relaxation.  In this configuration the spin density is distributed broadly throughout the lattice, and no single oxygen atom carries a large magnetic moment.  This indicates that the hole remains electronically delocalized in the absence of lattice relaxation.  The localized polaron state only forms after structural relaxation of the charged supercell indicating the known artifact of hybrid functionals to trap holes even in the absence of relaxation did not occur.  

To evaluate the thermodynamic stability of substitutional impurities and compensating intrinsic defects in $\mathrm{\kappa-Ga_2O_3}$, defect formation energies were calculated as a function of both the oxygen chemical potential \cite{lyons2018survey} and the Fermi level \cite{freysoldt2014first}.  As an example, the formation energy of the substitutional isoelectronic impurity $\mathrm{Al_{Ga}}$ in charge state q is given by

\begin{equation}
\label{eq2}
\mathrm{E^f[Al_{Ga}^q]=E_{tot}[Al_{Ga}^q]-E_{tot}[Ga_2O_3]+\mu_{Ga}-\mu_{Al}+qE_F+E_{corr}}
\end{equation}

where $\mathrm{E_{tot} [Al_{Ga}^q ]}$ is the total energy of the supercell containing an Al impurity in charge state q and $\mathrm{E_{tot} [Ga_2 O_3 ]}$ is the total energy of pristine $\mathrm{\kappa-Ga_2O_3}$ in the same supercell.  Electron exchange with the electron reservoir is accounted for by the Fermi level $\mathrm{E_F}$ relative to the VBM and $\mathrm{E_{corr}}$ is the finite size correction   term   to account for interactions between mirror charges in repeat cells \cite{kumagai2014electrostatics}.

Removing a Ga atom causes exchange with a reservoir of energy $\mathrm{\mu_{Ga}}$ which is referenced to the energy per atom of metallic Ga in the 8-atom orthorhombic structure.  The range of values for the Ga reservoir energy are limited to the enthalpy of formation of $\mathrm{\kappa-Ga_2O_3}$ calculated to be $\mathrm{\Delta H_f(Ga_2 O_3 )=-12.42 eV}$.   Oxygen is referenced to half the energy of an $\mathrm{O_2}$ molecule.  The formation enthalpy of each stable binary oxide is used to limit the chemical potential range for each substitutional impurity.  For $\mathrm{X=Al,  In}$, the range of chemical potential is limited by $\mathrm{2\mu_X + 3\mu_O < \Delta H_f(X_2 O_3)}$, where the calculated enthalpies were $\mathrm{\Delta H_f (Al_2 O_3)=-18.64 eV}$ and $\mathrm{\Delta H_f(In_2 O_3 )=-10.76 eV}$.  For $\mathrm{X=Mg, Zn}$, the range is limited by $\mathrm{\mu_X + \mu_O < \Delta H_f (XO)}$, with calculated enthalpies of $\mathrm{\Delta H_f(MgO)=-6.57 eV}$ and $\mathrm{\Delta H_f(ZnO)=-3.91 eV}$.  

Finite-size corrections for charged defects were applied using the anisotropic correction method of Kumagai and Oba \cite{kumagai2014electrostatics} with a calculated anisotropic dielectric function of $(\mathrm{\epsilon_{xx},\epsilon_{yy},\epsilon_{zz}}) = (20.58, 17.79, 19.57)$.  The dielectric function was computed using density functional perturbation theory.  Formation energies were evaluated as a function of the Fermi level, varied from the valence band maximum (VBM) which is set to 0 eV, to the CBM in O-poor and O-rich conditions.  For cation substitution, each distinct coordination site was examined, and the lowest-energy configuration was adopted for each impurity; the same procedure was followed for oxygen vacancies. Hole self-trapping energies were determined for each lowest-energy impurity configuration, while formation energies were calculated for all lowest-energy defect structures using the standard formalism of Freysoldt et al \cite{freysoldt2014first}.	

\section{\label{sec:results}Results and Discussion}
Removal of an electron from the neutral supercell results in localization of the hole on an oxygen atom, forming a small polaron accompanied by local lattice relaxation. The calculated spin density is strongly localized on a single oxygen site, consistent with the expected behavior for oxides in which the valence band is dominated by O 2p states \cite{lyons2014effects}.  Structural relaxation of the charged supercell produces a distortion of the surrounding Ga–O bonds that stabilizes the localized hole state. The Ga atoms coordinated to the hole-hosting oxygen relax outward relative to the neutral structure, yielding a maximum structural distortion of $\mathrm{\Delta R_{max} = 0.044 Å}$. This lattice relaxation is characteristic of small hole polarons in oxide materials, where localization of the hole is stabilized by expansion of the surrounding cation–oxygen bonds.  

For $\mathrm{\kappa-Ga_2O_3}$ the calculated hole self trapping value is $\mathrm{E_{ST} = 0.173 eV}$, indicating that hole localization is energetically favored relative to the delocalized state.  Hole self-trapping has previously been reported in the $\beta$ phase of $\mathrm{Ga_2O_3}$, where hybrid functional calculations similarly predict localization of the hole on an oxygen atom accompanied by outward relaxation of neighboring Ga–O bonds. Reported values of $\mathrm{E_{ST}}$ for $\mathrm{\beta-Ga_2O_3}$ are typically on the order of ~0.1–0.2 eV, depending on the computational approach and the specific oxygen coordination environment [1,4,5]. The present results therefore indicate that $\mathrm{\kappa-Ga_2O_3}$ exhibits hole localization behavior comparable to that observed in the $\beta$ phase, confirming that small hole polarons are an intrinsic feature of $\mathrm{Ga_2O_3}$ polymorphs.

Previous hybrid functional calculations have also examined hole self-trapping in $\mathrm{\kappa-Ga_2O_3}$.  Prior studies reported a larger self-trapping energy for the $\kappa$ phase using a similar HSE-based approach.  The somewhat smaller value obtained in the present work likely arises from several methodological differences, including the use of a denser k-point sampling of the Brillouin zone and the inclusion of Ga 3d electrons in the valence.  In addition, the present calculations employ structural distortion searches to more thoroughly sample the potential energy surface prior to hybrid-functional relaxation \cite{mosquera2023identifying}.  These differences can influence both the delocalized reference state and the relaxed polaron configuration, which leads to modest variations in the calculated self-trapping energy.  Despite these differences, both studies consistently predict that $\mathrm{\kappa-Ga_2O_3}$ supports a stable self-trapped hole.  Establishing the intrinsic behavior of the self-trapped hole provides a reference point for evaluating how substitutional impurities modify the stability and structure of the localized hole state in $\mathrm{\kappa-Ga_2O_3}$.

To examine how substitutional impurities modify hole localization in $\mathrm{\kappa-Ga_2O_3}$, the self-trapping energies and associated lattice distortions were calculated for Al, In, Mg, and Zn substitution on the Ga site.  Al, Zn, and Mg favor substitution at the tetrahedral site (I) while In prefers the octahedral site (II); the O vacancy prefers the 3-fold Ga-coordinated position.  The resulting self-trapping energies ($\mathrm{E_ST}$) and maximum structural distortions ($\mathrm{\Delta R_{max}}$) are summarized in Table \ref{tab:table1}.  These distortions are a measure of the bond changes in the cell between the neutral system and the same system with a hole, maintaining the same lattice constants and allowing for full ionic relaxation.  

\begin{table}[!ht]
    \centering
    \caption{Self-trapping energies $\mathrm{E_{ST}}$ and maximum lattice distortions $\mathrm{\Delta R_{max}}$ associated with hole localization in $\mathrm{\kappa-Ga_2O_3}$ and in $\mathrm{\kappa-Ga_2O_3}$ containing substitutional impurities on the Ga site.}
    \begin{tabular}{|c|c|c|}
        \hline System & $\mathrm{E_ST (eV)}$ & $\mathrm{\Delta R_{max} (Å)}$ \\
        \hline $\mathrm{\kappa-Ga_2O_3}$ & 0.173 & 0.044 \\
        \hline $\mathrm{In_{Ga}}$ & -0.200 & 0.020 \\
        \hline $\mathrm{Mg_{Ga}}$ & 0.131 & 0.037 \\
        \hline $\mathrm{Al_{Ga}}$ & -0.215 & 0.014 \\
        \hline $\mathrm{Zn_{Ga}}$ & 0.453 & 0.221 \\
        \hline
    \end{tabular}
    \label{tab:table1}
\end{table}

The influence of isoelectronic substitution on hole localization was examined for Al and In substitution on the Ga site. Because these impurities do not introduce free charge carriers, their primary effect is to modify the local bonding environment in which the hole polaron forms.  For both $\mathrm{Al_{Ga}}$ and $\mathrm{In_{Ga}}$, the calculated self-trapping energies are negative, indicating that holes favor delocalization.  The calculated values of -0.215 eV for $\mathrm{Al_{Ga}}$ and -0.200 eV for $\mathrm{In_{Ga}}$ are coupled with associated lattice distortions that are smaller than in intrinsic $\mathrm{\kappa-Ga_2O_3}$.  The maximum distortions are 0.014 Å for Al substitution and 0.020 Å for In substitution, both substantially smaller than the distortion observed for the intrinsic self-trapped hole.  This indicates that isoelectronic substitution in $\mathrm{\kappa-Ga_2O_3}$ modifies the bonds around the STH in the host material and delocalizes the polaron.  

The suppression of hole localization in the presence of impurities is consistent with the general behavior of hole polarons in WBOs, where the stability of the localized state depends on the impurity atoms present \cite{lyons2022self,varley2012role}.  A similar dependence of polaron stability on the local cation environment has been reported in studies of other $\mathrm{Ga_2O_3}$ polymorphs. First-principles calculations have shown that the magnitude of hole self-trapping energies varies across the different $\mathrm{Ga_2O_3}$ phases due to differences in local bonding geometry and lattice relaxation \cite{lyons2019electronic,gake2019first}. In the $\beta$ phase of $\mathrm{Ga_2O_3}$, substitutional impurities on the Ga site also influence the stability of localized hole states through changes in the surrounding lattice environment \cite{lyons2014effects}. 

\begin{figure}[!ht]
    \includegraphics[scale=0.5]{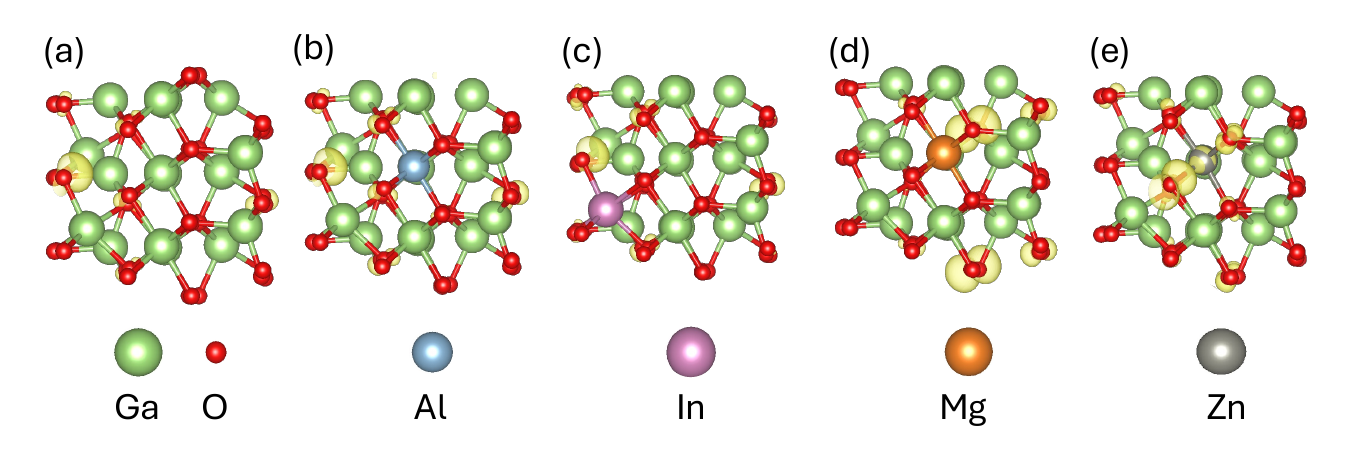}
    \caption{Spin density of the localized hole in (a) intrinsic $\mathrm{\kappa-Ga_2O_3}$, (b) $\mathrm{Al_{Ga}}$, (c) $\mathrm{In_{Ga}}$, (d) $\mathrm{Mg_{Ga}}$, and (e) $\mathrm{Zn_{Ga}}$.  Yellow isosurfaces represent the hole spin density set to $\mathrm{0.007 e/Å^3}$.  Red, green, blue, purple, orange, and silver atoms are O, Ga, Al, In, Mg, and Zn, respectively.}
    \label{fig:fig1}
\end{figure}

The spatial distribution of the localized hole state and its modification by substitutional impurities is shown in Figure \ref{fig:fig1}.  Yellow isosurfaces represent the spin density projected onto the relaxed structure and reveal the dominant O 2p orbital shape in each system; some spin density is 4s orbital-like in the Zn system.  

Figure \ref{fig:fig1}(a) shows the localized hole and accompanying lattice distortions in intrinsic $\mathrm{\kappa-Ga_2O_3}$.  The influence of substitutional impurities on the localized hole state is shown in Figure \ref{fig:fig1}(b-e).  The distortions are reduced for the isoelectronic substitutes $\mathrm{Al_{Ga}}$ and $\mathrm{In_{Ga}}$ resulting in a delocalization of the STH.  No oxygen atom possesses a spin density over $40\%$ of the total.

For $\mathrm{Mg_{Ga}}$, the spin density remains localized on an oxygen atom (over $70\%$ on a single O atom) and exhibits an orbital shape nearly identical to that observed in the intrinsic system.  This indicates that Mg does not fundamentally modify the character of the hole polaron and instead acts as a weak perturbation to intrinsic localization much like in the host material.  Similar behavior has been reported for acceptor impurities in $\mathrm{\beta-Ga_2O_3}$, where holes localize on nearby oxygen atoms rather than on the impurity itself \cite{lyons2022self,lyons2014effects}.  In those systems, the localized hole forms an oxygen-centered polaron that is weakly influenced by the acceptor impurity, consistent with the behavior observed here for Mg in $\mathrm{\kappa-Ga_2O_3}$.  The persistence of an oxygen-centered polaron also agrees with previous theoretical studies of self-trapped holes in $\mathrm{Ga_2O_3}$ polymorphs, which show that hole localization is largely governed by the O 2p character of the valence band \cite{lyons2019electronic,lyons2022self,varley2012role,gake2019first}. 

In contrast, $\mathrm{Zn_{Ga}}$ produces a noticeable modification of the localized hole state.  As shown in Figure \ref{fig:fig1}(e), the spin density remains primarily localized on oxygen atoms coordinated to the Zn impurity but exhibits hybridization between the O 2p and the Zn 4s orbitals.  The resulting distortion indicates the formation of an impurity-bound polaron complex rather than a purely intrinsic oxygen polaron. This behavior is consistent with previous studies of acceptor impurities in WBOs, where localized holes can couple strongly to specific impurities through hybridization \cite{lyons2022self,lyons2014effects}.  The presence of Zn–O hybridization is also consistent with earlier investigations of Zn incorporation in $\mathrm{\beta-Ga_2O_3}$ where significant modification of the electronic structure and defect states associated with Zn substitution were seen \cite{li2012electronic}.  The stronger interaction between Zn and the localized hole is reflected in the larger structural distortions and higher self-trapping energy obtained for $\mathrm{Zn_{Ga}}$, indicating that Zn stabilizes the localized hole state relative to both the intrinsic system and the Mg acceptor case.  This is likely due to the similar valence structure between Zn and Ga where Ga possesses one more electron in a 4p orbital.  

Chemical potentials were calculated at the O-rich and O-poor limits and are summarized in Table \ref{tab:table2}.   Values that did not satisfy the formation enthalpy limits were set equal to zero, except in the case of O-rich O and O-poor Ga, where their chemical potentials are defined to be zero such that the elemental chemical potential and energy of the stable monoatomic structure are equal i.e., $\mathrm{\Delta \mu_{i} = 0}$ for defect \emph{i}.  

\begin{table}[!ht]
    \centering
    \caption{Chemical potentials used in formation energy calculations, at the O-rich and O-poor limits.}
    \begin{tabular}{|c|c|c|}
        \hline Element & O-rich & O-poor \\
        \hline $\mathrm{\mu_{Ga}}$ & -6.208 & 0.000 \\
        \hline $\mathrm{\mu_{O}}$ & 0.000 & -4.138 \\
        \hline $\mathrm{\mu_{Al}}$ & -9.320 & -3.113 \\
        \hline $\mathrm{\mu_{In}}$ & -5.380 & 0.000 \\
        \hline $\mathrm{\mu_{Mg}}$ & -6.566 & -2.428 \\
        \hline $\mathrm{\mu_{Zn}}$ & -3.914 & 0.000 \\
        \hline
    \end{tabular}
    \label{tab:table2}
\end{table}

\begin{figure}[!ht]
    \includegraphics[scale=1]{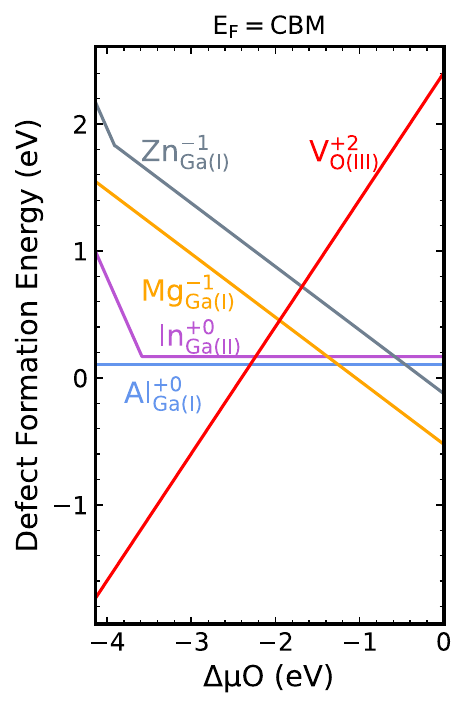}
    \caption{Formation energy of cation impurities and oxygen vacancies as a function of oxygen chemical potential.}
    \label{fig:fig2}
\end{figure}

The formation energies of substitutional impurities on the Ga site and oxygen vacancies are shown in Figure \ref{fig:fig2} as a function of the oxygen chemical potential.  Formation energies were evaluated with the Fermi level fixed at the CBM to represent the n-type limit  which reveals the stability of each impurity.  Rearranging Equation \ref{eq2} such that the oxygen chemical potential becomes the independent variable allows one to determine the stability of each impurity at the different growth limits.  At the O-rich condition ($\mathrm{\Delta \mu_O = 0 eV}$) the acceptor impurities form most readily and the oxygen vacancy is least likely to form.  At the O-poor end ($\mathrm{\Delta \mu_O = -4.138 eV}$) the O vacancy is most likely to form, suggesting compensation of any effects due to impurities at the cation site.  The inflection in the In and Zn systems is due to the chemical potential going to 0 eV such that the formation enthalpy is less than that of the host.    

The behavior of oxygen vacancies in the $\kappa$-phase is consistent with previous theoretical studies of $\mathrm{Ga_2O_3}$, which identify oxygen vacancies as dominant donor defects that limit the formation of free holes in WBOs \cite{lyons2022self,lyons2014effects}.  Previous theoretical studies of $\mathrm{\kappa-Ga_2O_3}$ have similarly identified oxygen vacancies as low-energy donor defects that can influence charge compensation in this polymorph \cite{lyons2019electronic,gake2019first}.  The present results are therefore consistent with earlier calculations indicating that intrinsic donor defects compete with acceptor impurities under oxygen-poor conditions.  The relative stability of $\mathrm{Mg_{Ga}}$ and $\mathrm{Zn_{Ga}}$ observed here further supports their consideration as candidate acceptor impurities in $\mathrm{\kappa-Ga_2O_3}$.  Similar trends have been reported previously for $\mathrm{\beta-Ga_2O_3}$.  

Hybrid functional calculations showed that Mg and Zn substitution on the Ga site are the most stable acceptor impurities in the $\beta$ phase, while oxygen vacancies act as donor defects under O-poor conditions \cite{lyons2019electronic,lyons2018survey}.  These vacancies compensate holes and therefore limit the effectiveness of p-type doping in $\mathrm{\beta-Ga_2O_3}$.  The behavior observed here for $\mathrm{\kappa-Ga_2O_3}$ follows the same general trend, indicating that compensation by oxygen vacancies remains an important limitation for acceptor doping across different $\mathrm{Ga_2O_3}$ polymorphs.

\begin{figure}[!ht]
    \includegraphics[scale=1]{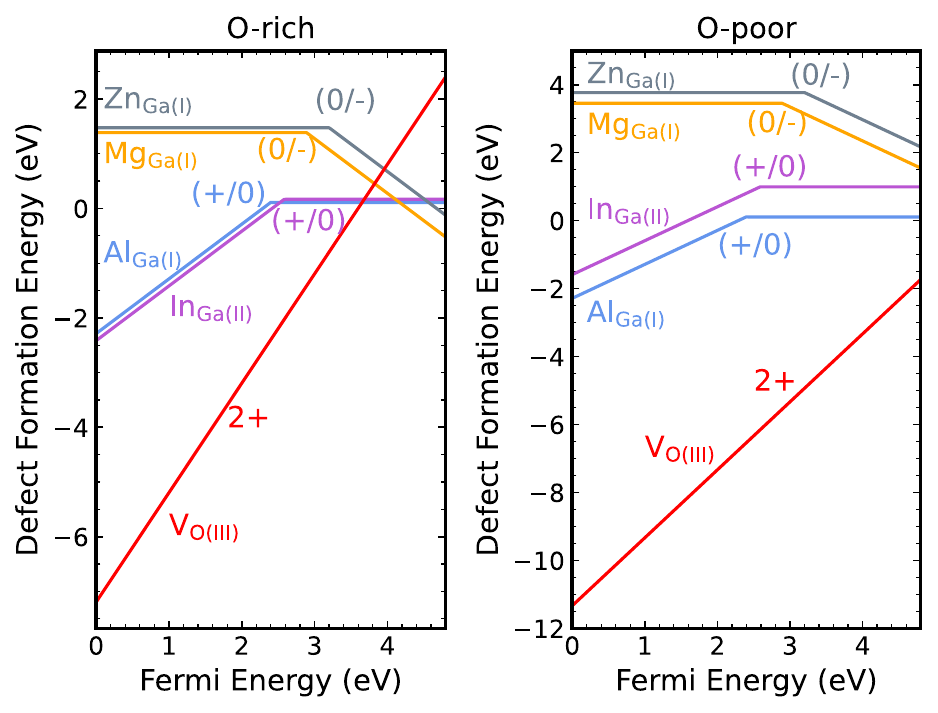}
    \caption{Formation energy in (a) O-rich and (b) O-poor conditions for cation impurities and oxygen vacancies as a function of the Fermi level as referenced from the maximum valence band energy which is set to 0 eV.  Charge transition states (q/q’) are noted as well as the favorable substitution site with the subscripts for each impurity (Ga(I) = tetrahedral coordination, Ga(II) = octahedral coordination, O(III) = 3-fold Ga coordination).}
    \label{fig:fig3}
\end{figure}

The dependence of defect formation energies on the Fermi level under O-poor and O-rich conditions is shown in Figure \ref{fig:fig3}.  Under O-rich conditions the lowest energy impurities show alternating behavior when traversing the bandgap energies.  Near the valence band edge O vacancies are most favorable while acceptor dopants are least, and the exact opposite is true at the conduction band edge.  In an O-poor environment, O vacancies are again the lowest energy at the valence band edge, but the trend is maintained across the bandgap i.e., acceptors are highest across the gap, isoelectronic impurities are in the middle and O vacancies are lowest energy.  In both O-rich and O-poor conditions, all electronic transitions are deep, occurring at least 2 eV above the VBM and about the same below the CBM i.e., transitions occur in the 2-3 eV range.  O vacancies only exist in the 2+ charge state across the entire band gap and are the lowest energy defects near the VBM.  Compensation due to O vacancies has been widely identified as a key limitation for achieving p-type conductivity in WBOs, including $\mathrm{Ga_2O_3}$ \cite{lyons2022self,lyons2014effects}.

These formation energy calculations provide important context for the polaron behavior relative to each impurity.  Near the VBM, isoelectronic impurities readily substitute on the Ga site but the transition representing the Fermi level at which an electron is removed to form a hole (+/0) is deep for both Al and In (above 2 eV for both).  This suggests that isoelectronic substitution destabilizes localized hole formation in the host material.  For the acceptor impurities, their substitution is more favorable in an O-rich environment and both Mg and Zn have ionization energies (0/-) over 3 eV.  This means that acceptor impurities tend to stabilize the STH as a deep acceptor level in the bandgap, a behavior seen across the polymorphs of $\mathrm{Ga_2O_3}$ for Mg \cite{lyons2019electronic}.

\section{\label{sec:con}Conclusions}
In summary, hybrid density functional theory calculations were used to investigate hole self trapping and the influence of substitutional impurities in $\mathrm{\kappa-Ga_2O_3}$.  In the intrinsic material, a hole localizes on an oxygen atom, forming a small polaron stabilized by local lattice relaxation.  Isoelectronic substitution by Al and In suppresses hole localization by reducing lattice distortions and destabilizing the polaron state.  Mg acts as a polaronic acceptor with a STH on a nearest-neighbor O and Zn introduces hybridization between the localized hole and the impurity site.  Formation energy calculations indicate that AlGa and InGa are thermodynamically accessible impurities, although compensation by oxygen vacancies becomes increasingly favorable.  These results illustrate how substitutional impurities perturb localized hole states in $\mathrm{\kappa-Ga_2O_3}$ and highlight the interplay between polaron formation and defect thermodynamics in determining the electronic behavior of WBOs like the $\kappa$-phase of $\mathrm{Ga_2O_3}$.

\section{Acknowledgments}
EW would like to acknowledge fruitful conversations with Dr. Jack (John) Lyons and Dr. Seán Kavanagh.  EW and LS would also like to acknowledge help in data procurement from Mr. Lauro Guerra and for discussions about devices with experimentalists from the group of Dr. Ravi Droopad at Texas State University. We acknowledge the use of the high-performance computing (LEAP2) center at Texas State University.  The work was partially supported by the Office of Naval Research Grant Number W911NF-25-1-0107 and the National Science Foundation Award Number: 2514718 (CREST Phase I Center for Ultrawide Bandgap Semiconductor Devices Materials).  The numpy\cite{harris2020array}, pymatgen\cite{jain2011high}, scipy\cite{virtanen2020scipy}, matplotlib\cite{hunter2007matplotlib}, doped\cite{kavanagh2024doped}, shakenbreak\cite{mosquera2022shakenbreak}, sumo\cite{ganose2018sumo}, and spinney\cite{arrigoni2021spinney} python libraries and the vaspkit\cite{wang2021vaspkit} suites were used to create inputs, parse outputs, and post-process and plot data.  The scripts used and data generated in this study may be obtained from the authors upon reasonable request.  

\bibliography{bibtex}

\end{document}